\documentclass[prd,twocolumn,showpacs,preprintnumbers,amsmath,amssymb,superscriptaddress,floatfix,nofootinbib]{revtex4-2}

\usepackage{graphicx}
\usepackage{bm}
\usepackage{amsmath}
\usepackage{amsfonts}
\usepackage{amssymb}
\usepackage{color}
\usepackage{multirow}
\usepackage{subfigure}
\usepackage[colorlinks, citecolor=blue,anchorcolor=red,menucolor=red, linkcolor=red,filecolor=red,runcolor=red,urlcolor=blue,frenchlinks=red]{hyperref}

\begin{document}
\title{Theoretical study of  the open-flavored tetraquark $T_{c\bar{s}}(2900)$ in the process $\Lambda_b\to K^0D^0\Lambda$}

\author{Wen-Tao Lyu}
\affiliation{School of Physics, Zhengzhou University, Zhengzhou 450001, China}

\author{Yun-He Lyu}
\affiliation{School of Physics, Zhengzhou University, Zhengzhou 450001, China}\vspace{0.5cm}

\author{Man-Yu Duan}
\affiliation{School of Physics, Southeast University, Nanjing 210094, China}

\author{Guan-Ying Wang}
\email{wangguanying@henu.edu.cn}
\affiliation{Joint Research Center for Theoretical Physics, School of Physics and Electronics, Henan University, Kaifeng 475004, China}\vspace{0.5cm}

\author{Dian-Yong Chen}\email{chendy@seu.edu.cn}
\affiliation{School of Physics, Southeast University, Nanjing 210094, China}
\affiliation{Lanzhou Center for Theoretical Physics, Lanzhou University, Lanzhou 730000, P. R. China}

\author{En Wang}
\email{wangen@zzu.edu.cn}
\affiliation{School of Physics, Zhengzhou University, Zhengzhou 450001, China}\vspace{0.5cm}

\begin{abstract}

Recently, the LHCb Collaboration has measured the processes $B^0\to\bar{D}^0D_s^+\pi^-$ and $B^+\to D^-D_s^+\pi^+$, where the $D_s^+\pi^-$ and $D_s^+\pi^+$ invariant mass distributions show the significant signals of two new open-flavored tetraquark states $T_{c\bar{s}}(2900)^0$ and $T_{c\bar{s}}(2900)^{++}$, as the two of the isospin triplet. In this work, we have investigated the process $\Lambda_b\to K^0D^0\Lambda$ by taking into account  the tetraquark state $T_{c\bar{s}}(2900)^0$ and the intermediate nucleon resonance  $N^*(1535)$, which could be dynamically generated by the interactions of the $D^*K^*/D^*_s\rho$ and the pseoduscalar meson-octet baryon, respectively. Our results show that a clear peak of the open-flavored tetraquark $T_{c\bar{s}}(2900)$ may appear in the $K^0D^0$ invariant mass distribution of the process $\Lambda_b\to K^0D^0\Lambda$, which could be tested by future experiments.
\end{abstract}

\date{\today}

\maketitle

\section{Introduction}\label{sec1}

In 2020, the LHCb Collaboration has analyzed the  relevant data of the process $B^+\to D^+D^-K^+$, and observed two resonances $X_0(2900)$ and $X_1(2900)$ with masses about 2900~MeV in the $D^-K^+$ invariant mass distribution~\cite{LHCb:2020bls,LHCb:2020pxc}. These two resonances have attracted lots of attention due to the fact that they contain four different flavor quarks ($\bar{c}\bar{s}ud$) and can not be described within the conventional quark models. Some exotic interpretations have been proposed, such as compact tetraquark, molecular states, and sum rule derivations,  (among other see Ref.\cite{Dai:2022qwh}). Since the  mass of the $X_0(2900)$ is close to the $\bar{D}^*K^*$ threshold, 
the $\bar{D}^*K^*$ molecular explanation is more popular~\cite{Liu:2020nil, Huang:2020ptc,Hu:2020mxp,Kong:2021ohg,Wang:2021lwy,Xiao:2020ltm,Lin:2022eau,Chen:2020eyu,Liu:2020orv, Chen:2020aos,Chen:2021erj,Albuquerque:2020ugi,Xiao:2020ltm}.
In particular, by using the method of QCD sum rules, the authors in Refs.~\cite{Chen:2020aos,Chen:2021erj,Albuquerque:2020ugi} have interpreted the $X_0(2900)$ as the $S$-wave $\bar{D}^{*}K^{*}$ molecular state with $J^P=0^+$, and suggested the explanation of $X_1(2900)$ as the $P$-wave $\bar{c}\bar{s}ud$ compact tetraquark state with $J^P=1^-$. The estimations within the one-boson exchange model indicated that the $X_0(2900)$ could be a $\bar{D}^*K^*$ molecule state~\cite{Liu:2020nil}.
 
Two years after the observation of $X_{0,1}(2900)$, another two tetraquark candidates with charmness and strangeness around 2900 MeV, were reported by the LHCb Collaboration in the $D_s^+\pi^-$ and $D_s^+\pi^+$ invariant mass distributions of the $B^0\to\bar{D}^0D_s^+\pi^-$ and $B^+\to D^-D_s^+\pi^+$ decays, respectively~\cite{LHCb:2022sfr,LHCb:2022lzp}. The quark components of these two states should be $c\bar{s}\bar{u}d$ and $c\bar{s}u\bar{d}$, and their masses and widths are measured to be,
\begin{equation}
	\begin{aligned}
		& M_{T_{c \bar{s}}(2900)^0}=(2892 \pm 14 \pm 15) \mathrm{MeV} \text { , } \\
		& \Gamma_{T_{c \bar{s}}(2900)^0}=(119 \pm 26 \pm 12) \mathrm{MeV} \text { , } \\
		& M_{T_{c \bar{s}}(2900)^{++}}=(2921 \pm 17 \pm 19) \mathrm{MeV} \text { , } \\
		& \Gamma_{T_{c \bar{s}}(2900)^{++}}=(137 \pm 32 \pm 14) \mathrm{MeV} \text { . } 
	\end{aligned}
\end{equation}
The significance is found to be $8.0\sigma$ for the $T_{c\bar{s}}(2900)^0$ state and $6.5\sigma$ for the $T_{c\bar{s}}(2900)^{++}$ state. Taking into account that their masses and widths are close to each other within the uncertainties,  $T_{c\bar{s}}(2900)^{++}$ and $T_{c\bar{s}}(2900)^0$ should be two of the isospin triplet.

It is worth to mention that such kind of states were predicted more than ten years ago in Ref.~\cite{Molina:2010tx}. After the observation of $T_{c\bar{s}}(2900)^{++}$ and $T_{c\bar{s}}(2900)^0$, there are several interpretations about their structure. The proximity of the $D^*K^*$ and the $D_s^*\rho$ thresholds to the mass of $T_{c\bar{s}}(2900)$ suggests that these two-hadron channels could play important roles in the dynamics of the $T_{c\bar{s}}(2900)$ states, hinting to a hadronic molecular interpretation of the two states~\cite{Yue:2022mnf,Chen:2022svh}. For instance, Ref.~\cite{Agaev:2022duz} argues that $T_{c\bar{s}}(2900)^{++}$ and $T_{c\bar{s}}(2900)^0$ may be modeled as molecules $D_s^{*+}\rho^+$ and $D_s^{*+}\rho^-$, respectively, using two-point sum rule method.  Recently, within the local hidden symmetry formalism,  the $T_{c\bar{s}}(2900)$ could be explained as a $D^*_s\rho-D^*K^*$ bound/virtual state in Ref.~\cite{Duan:2023lcj}.
 And we have proposed to search for the open-flavored tetraquark $T_{c\bar{s}}(2900)$ in the processes $B^+\to D^+D^-K^+$~\cite{Duan:2023qsg}, $\bar{B}_s^0\to K^0D^0\pi^0$~\cite{Lyu:2024wxa}, and $B^-\to D_s^+K^-\pi^-$~\cite{Lyu:2023jos} by assuming $T_{c\bar{s}}(2900)$ as a $D^*K^*$ molecular state. Meanwhile, its spin partner with $J^P=2^+$ predicted in Refs.~\cite{Molina:2022jcd,Duan:2023lcj} is suggested to be studied in the processes $B\to D^{*-}D^+K^+$ and  $\Lambda_b\to \Sigma_c^{++}D^{-}K^{-}$~\cite{Lyu:2024zdo,Song:2024dan}.

In addition, the compact tetraquark interpretation of  the $T_{c\bar{s}}(2900)$ with quark contents $[c\bar{s}\bar{u}d]$ and $[c\bar{s}u\bar{d}]$ is studied in Refs.~\cite{Yang:2023evp, Lian:2023cgs, Jiang:2023rcn, Liu:2022hbk, Dmitrasinovic:2023eei, Ortega:2023azl}, while the $T_{c\bar{s}}(2900)$ structure is also proposed to be a threshold cusp effect from the interaction between the $D^*K^*$ and $D^*_s\rho$ channels~\cite{Molina:2022jcd} or the kinetic effect from a triangle singularity~\cite{Ge:2022dsp}. Thus, searching for the $T_{c\bar{s}}(2900)$ in other processes is crucial to exclude the interpretation of the kinetic effects, and the more experimental information could be helpful to explore its structure and deepen our understanding of the hadron-hadron interactions.

The $\Lambda_b$ decays to heavy particles with charmness show an immense potential to the investigation of particle properties~\cite{Fayyazuddin:2017loe,Wei:2021usz,Liu:2020ajv,Wang:2015pcn,Lu:2016roh}. Taking into account that the $T_{c\bar{s}}(2900)$ is predicted to couple strongly to the $DK$ channel~\cite{Duan:2023lcj,Yue:2022mnf,Lian:2023cgs}, we would like to propose to search for this state in the $\Lambda_b\to D^0K^0\Lambda$, which has not yet been measured experimentally. However, the process $\Lambda_b\to D^0p\pi^-$ has been observed by LHCb Collaboration~\cite{LHCb:2013jih,LHCb:2017jym}.
With the ratio of the branching fractions measured by LHCb,
\begin{eqnarray}
R_{\Lambda_b\to D^0p\pi^-}&=&\frac{\mathcal{B}(\Lambda_b\to D^0p\pi^-)\times \mathcal{B}(D^0\to K^-\pi^+)}{\mathcal{B}(\Lambda_b\to \Lambda_c\pi^-)\times \mathcal{B}(\Lambda_c\to pK^-\pi^+)} \nonumber \\
&=&0.0806\pm0.0023\pm0.0035,   \nonumber 
\end{eqnarray}
and the  data of Review of Particle Physics (RPP)~\cite{ParticleDataGroup:2022pth},
\begin{eqnarray}
 \mathcal{B}(D^0\to K^-\pi^+)&=& (3.947\pm 0.030)\%, \nonumber \\ 
 \mathcal{B}(\Lambda_b\to \Lambda_c\pi^-)&=&(4.9\pm0.4)\times 10^{-3}, \nonumber \\ 
 \mathcal{B}(\Lambda_c\to pK^-\pi^+)&=& (6.28\pm 0.32)\%, \nonumber 
\end{eqnarray}
we could roughly extract the branching fraction $\mathcal{B}(\Lambda_b\to D^0p\pi^-)=6.28\times 10^{-4}$ by applying the narrow width approximation~\cite{Cheng:2020iwk}.  Since both the tree-diagram amplitudes of the processes $\Lambda_b\to D^0K^0\Lambda$ and $\Lambda_b\to D^0p\pi^-$ are in proportion to the Cabbibo-Kobayashi-Maskawa (CKM) matrix elements $V_{cb}V_{ud}$, it is expected that the branching fractions of those processes should be of the same order of magnitude, thus the process $\Lambda_b\to D^0K^0\Lambda$ is expected to be observed by LHCb Collaboration. Thus, in this work we would like to investigate the process $\Lambda_b\to D^0K^0\Lambda$ to show the possible evidence of the $T_{c\bar{s}}(2900)$ in the $D^0K^0$ invariant mass distribution. Here we focus on the process $\Lambda_b\to D^0K^0\Lambda$ decay by considering the $D_s^{*+}\rho^-$ and $D^{*0}K^{*0}$ final-state interaction to generate the signal of $T_{c\bar{s}}(2900)^0$. In addition, we will also consider the contribution from the $S$-wave pseudoscalar meson-octet baryon interaction within the unitary chiral approach, which could dynamical generate the state $N^*(1535)$~\cite{Inoue:2001ip}. 

This paper is organized as follows. In Sec.~\ref{sec2}, we present the theoretical formalism of the process $\Lambda_b\to D^0K^0\Lambda$. Numerical results are presented in Sec.~\ref{sec3}. Finally, we present a short summary in the last section.

 \section{Formalism}\label{sec2}
 
In this section, we will introduce the theoretical formalism for the process $\Lambda_b\to D^0K^0\Lambda$. 
Firstly, the mechanism for this process via the intermediate state $T_{c\bar{s}}(2900)$ is given in Sec.~\ref{sec2a}, and then the mechanism via the intermediate state $N^*(1535)$  is given in Sec.~\ref{sec2b}. Finally, we give the formalism for the invariant mass distributions for this process in Sec.~\ref{sec2e}.

 \subsection{$T_{c\bar{s}}(2900)$ in the $\Lambda_b\to D^0K^0\Lambda$} \label{sec2a}

\begin{figure}[htbp]
	\subfigure[]{\includegraphics[scale=0.63]{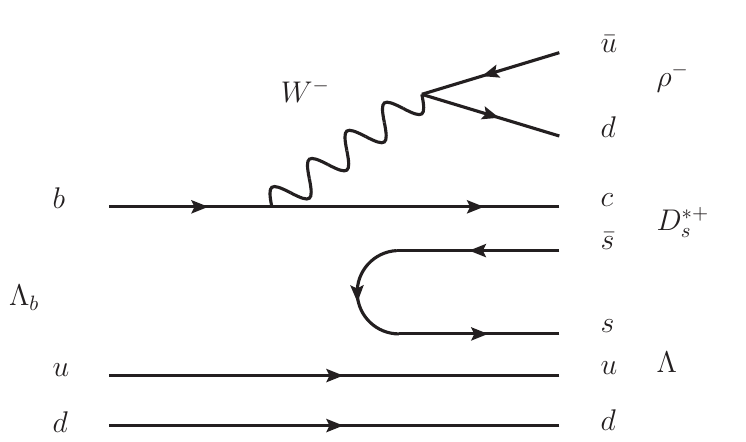}}	
	\subfigure[]{\includegraphics[scale=0.63]{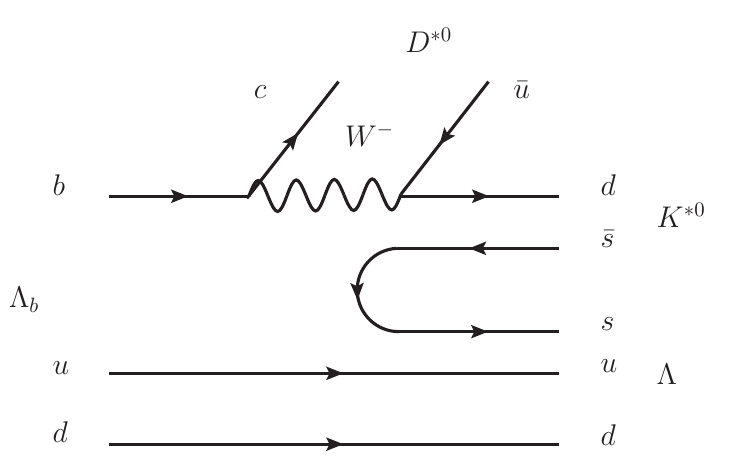}}
	\caption{The $\Lambda_b$ weakly decay at the microscopic quark picture. (a) The $\Lambda_b\to\rho^-D_s^{*+}\Lambda$ decay via the mechanism of the external $W^-$ emission, (b) the $\Lambda_b\to D^{*0}K^{*0}\Lambda$ decay via the mechanism of the internal $W^-$ emission.}\label{fig:Tcs-quark}
\end{figure}

Since the $T_{c\bar{s}}(2900)$ could be explained as the $D^*_s\rho-D^*K^*$ bound/virtual state in the coupled-channel approach~\cite{Duan:2023lcj}, one needs to produce the $D^*_s\rho$ or $D^*K^*$ via the mechanisms of the external $W^-$ emission and the internal $W^-$ emission, respectively as depicted in Figs.~\ref{fig:Tcs-quark}(a) and \ref{fig:Tcs-quark}(b).
 
For the mechanism of the external $W^-$ emission, the $b$ quark in the initial $\Lambda_b$ weakly transits into a $W^-$ boson and a $c$ quark, following by the $W^-$ decaying into the $\bar{u}d$ pair. The $ud$ pair of the initial $\Lambda_b$ has the isospin $I=0$, and acts as the spectator in this process. The $\bar{u}d$ pair from the $W^-$ boson decay will hadronize into $\rho^-$, and the $c$ and $ud$ pair, together with the $\bar{s}s$ pair created from the vacuum, will hadronize into the meson-baryon pairs as follows,
\begin{eqnarray}\label{H1}
	\left| \Lambda_b \right\rangle &=& \frac{1}{\sqrt{2}}b\left(ud-du\right) \nonumber \\
	&\Rightarrow & W^- c  \frac{1}{\sqrt{2}} \left(ud-du\right) \nonumber \\
	&\Rightarrow & \bar{u} dc \left(\bar{s}s\right) \frac{1}{\sqrt{2}} \left(ud-du\right) \nonumber \\
	&\Rightarrow & \rho^- c\left( \bar{s}s\right) \frac{1}{\sqrt{2}} \left(ud-du\right) \nonumber \\
	&\Rightarrow & \rho^- D_s^{*+} \frac{1}{\sqrt{2}} s\left(ud-du\right) \nonumber \\
	&\Rightarrow &-\sqrt{\frac{2}{3}} \rho^-D_s^{*+} \Lambda ,
\end{eqnarray}
where we use the meson flavor wave functions $\rho^-=\bar{u} d$, $D_s^{*+}=c\bar{s}$, and take the mixed
antisymmetric baryon wave functions as~\cite{Close:1979bt,Miyahara:2016yyh},
\begin{eqnarray}
\left| \Lambda_b \right>&=&\frac{1}{\sqrt{2}}\left[ b (ud-du) \right], \label{eq:wf_lambdab} \\
\left| \Lambda \right>&=& \frac{1}{\sqrt{12} }\left[u(ds-sd)+d(su-us)
-2s(ud-du)\right]. \label{eq:wf_lambda} \nonumber \\
\end{eqnarray}
Here it should be stressed that flavor wave function of $\Lambda$ is different with the one of Ref.~\cite{Close:1979bt}. By comparing the phase convention of Ref.~\cite{Close:1979bt} with the one inherent in the baryon octet matrix used in the chiral Lagrangians,
\begin{equation}
	{B}=\left(\begin{array}{ccc}
		\frac{\Sigma^0}{\sqrt{2}} +\frac{\Lambda}{\sqrt{6}} & \Sigma^{+} & p \\
		\Sigma^{-} & -\frac{\Sigma^0}{\sqrt{2}} +\frac{\Lambda}{\sqrt{6}}  & n \\
		\Xi^{-} & \Xi^0 & -\frac{2\Lambda}{\sqrt{6}} 
	\end{array}\right),
\end{equation}
one find that the phases of $\Sigma^+$, $\Lambda$, and $\Xi^0$ from Ref.~\cite{Close:1979bt} must be changed to agree with the Chiral Lagrangians, as discussed in Refs.~\cite{Pavao:2017cpt,Miyahara:2016yyh}.

On the other hand, we also can produce the $D^{*0} K^{*0}$ via the mechanism of the internal $W^-$ emission, as depicted in Fig.~\ref{fig:Tcs-quark}(b) as follows,
\begin{eqnarray}\label{H2}
	\left| \Lambda_b \right\rangle &=& \frac{1}{\sqrt{2}}b\left(ud-du\right) \nonumber \\
	&\Rightarrow & cW^- \frac{1}{\sqrt{2}} \left(ud-du\right) \nonumber \\
	&\Rightarrow & c\bar{u} d \left( \bar{s}s\right) \frac{1}{\sqrt{2}} \left(ud-du\right) \nonumber \\
	&\Rightarrow & -\sqrt{\frac{2}{3}} D^{*0} K^{*0} \Lambda.
\end{eqnarray}

		\begin{figure}[htbp]
	\subfigure[]{\includegraphics[scale=0.63]{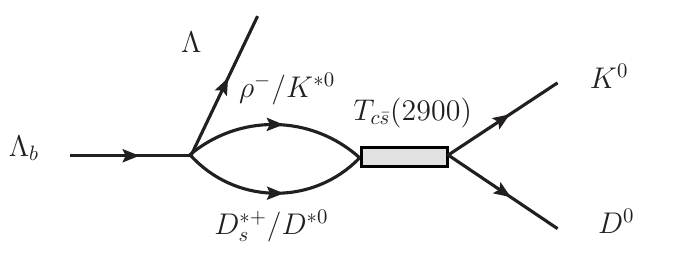}}	
	\subfigure[]{\includegraphics[scale=0.5]{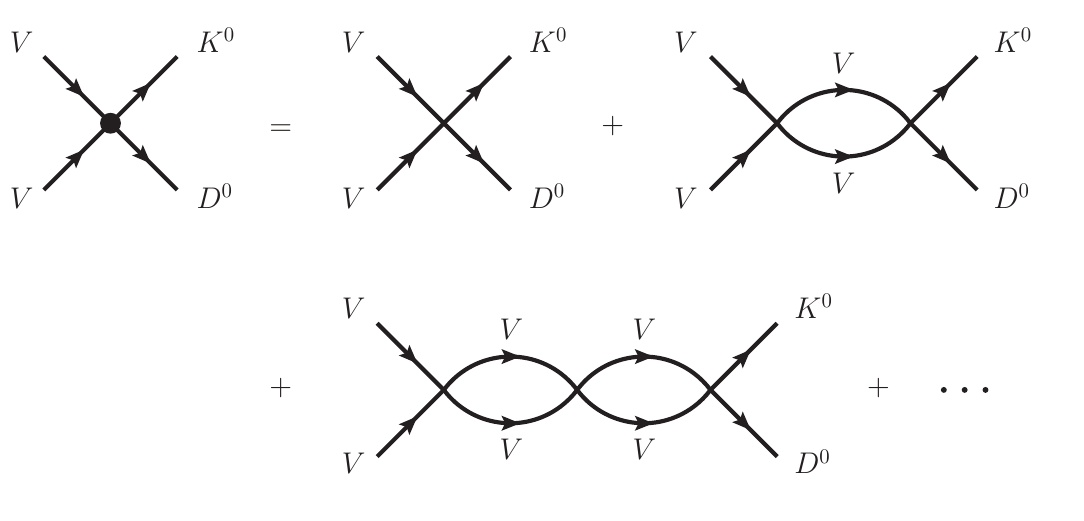}}
	\caption{(a) $\Lambda_b\to D^0K^0\Lambda$ decay via the intermediate state $T_{c\bar{s}}(2900)$ produced by the interactions of $D_s^{*+}\rho^-$ and $D^{*0}K^{*0}$. (b) The re-scattering process within a trivial summation.}\label{fig:Tcs-loop}
\end{figure}

Now, the states $D^{*+}_s\rho^-$
and $D^{*0} K^{*0}$ could undergo the  $S$-wave interactions to generate the $T_{c\bar{s}}(2900)$ state, following by decaying into the final states $D^0K^0$, as depicted in Fig.~\ref{fig:Tcs-loop}(a). 
In this case, we could obtain the unitary amplitude for the final state interactions following a trivial summation as depicted in Fig..~\ref{fig:Tcs-loop}(b), 
\begin{eqnarray}
 &&  \tilde{G}_{i} V'_{i\to D^0K^0} + \tilde{G}_{i} V_{i\to j} \tilde{G}_{j} V'_{j\to D^0K^0} +... \nonumber \\
 &=& \tilde{G}[1-V\tilde{G}]^{-1}V',
\end{eqnarray}
where $\tilde{G}$ is the loop functions of the two-meson system, $V'_{i\to D^0K^0}$ is the transition amplitude between the vector-vector channel and the $D^0K^0$ channel, and $V_{i\to j}$ are the transition potential between vector-vector channels~\cite{Molina:2022jcd},
\begin{equation}
			{V}=\left(\begin{array}{cc}
				0.11g^2 & -6.8g^2 \\
				-6.8g^2& 0 
			\end{array}\right),
\end{equation}
with $g=M_V/2f$, and $i,j=1,2$ corresponding to the $D_s^*\rho$, $D^{*0}K^{*0}$. Here we take the $M_V=M_\rho$ and  the pion decay constant $f=93$~MeV~\cite{Molina:2010tx}.

By defining the $\tilde{t}=[1-V\tilde{G}]^{-1}$, we could write the amplitude for Fig.~\ref{fig:Tcs-loop}(a),
   \begin{equation}\label{ttcs}
			\begin{aligned}
				\mathcal{T}^{T_{c\bar{s}}} = & -\sqrt{\frac{2}{3}}  V_p \left[C\times \tilde{G}_{D_s^{*+}\rho^-}\tilde{t}_{D_s^{*+}\rho^-\to D_s^{*+}\rho^-}V'_{D_s^{*+}\rho^-\to D^0K^0} \right.  \\
				& \left.+ C\times \tilde{G}_{D_s^{*+}\rho^-}\tilde{t}_{D_s^{*+}\rho^-\to D^{*0}K^{*0}}V'_{D^{*0}K^{*0}\to D^0K^0}\right. \\
				&\left.+ \tilde{G}_{D^{*0}K^{*0}}\tilde{t}_{D^{*0}K^{*0}\to D_s^{*+}\rho^-}V'_{D_s^{*+}\rho^-\to D^0K^0}\right. \\
				&\left.+ \tilde{G}_{D^{*0}K^{*0}}\tilde{t}_{D^{*0}K^{*0}\to D^{*0}K^{*0}}V'_{D^{*0}K^{*0}\to D^0K^0}\right],  
			\end{aligned}
		\end{equation}
	where the parameter $V_{p}$ contains all dynamical factors  of the production vertex of Fig.~\ref{fig:Tcs-quark}(b). Since in this work we mainly focus on the final state interactions of this process, and $\Lambda_b$ could decay into $\Lambda$ and $K^{*0} D^{*0}$ system with $J^P=0^+$ in $S$-wave, we assume $V_{p}$ to be constant, as done in Refs.~\cite{Wei:2021usz,Liu:2020ajv,Wang:2015pcn,Lu:2016roh}. 
 In addition, for the $W^-$ external emission of Fig.~\ref{fig:Tcs-quark}(a), the $\bar{u}d$ pair from the $W^-$ boson can form the color singlet $\rho^-$, thus the $\bar{u}$ and $d$ quarks could have three choices of colors. However, For the $W^-$ internal emission of Fig.~\ref{fig:Tcs-quark}(b), the $c$ quark from the $b$, with the same color as the $b$ quark, and the $\bar{u}$ quark with the anti-color will hadronize to the color singlet $D^{*0}$, which implies that all the quarks in the final states have the fixed colors.  Thus, we include a color factor $C=3$ to account for the relative weight of the $W^-$ external emission of Fig.~\ref{fig:Tcs-quark}(a) with respect to the $W^-$ internal mission of Fig.~\ref{fig:Tcs-quark}(b)  in the case of color number $N_C=3$~\cite{Duan:2020vye,Zhang:2020rqr,Dai:2018nmw}.  Within the dimensional regularization method, the loop function $G_{i}$ can be written as,
		\begin{equation}\label{Gb}
			\begin{aligned}
				G_{i} & =i \int \frac{d^4 q}{(2 \pi)^4} \frac{1}{(p-q)^2-m_2^2+i \epsilon} \frac{1}{q^2-m_1^2+i \epsilon} \\
				& =\frac{1}{16 \pi^2}\left\{a_i+\ln \frac{m_1^2}{\mu^2}+\frac{s+m_2^2-m_1^2}{2 s} \ln \frac{m_2^2}{m_1^2}\right. \\
				& +\frac{|\vec{q}\,|}{\sqrt{s}}\left[\ln \left(s-\left(m_2^2-m_1^2\right)+2 |\vec{q}\,| \sqrt{s}\right)\right. \\
				& +\ln \left(s+\left(m_2^2-m_1^2\right)+2 |\vec{q}\,| \sqrt{s}\right) \\
				& -\ln \left(-s+\left(m_2^2-m_1^2\right)+2 |\vec{q}\,| \sqrt{s}\right) \\
				& \left.\left.-\ln \left(-s-\left(m_2^2-m_1^2\right)+2 |\vec{q}\,| \sqrt{s}\right)\right]\right\},
			\end{aligned}
		\end{equation}
	where $m_{1,2}$ are the meson masses of the channels $D_s^{*+}\rho^-$ and $D^{*0}K^{*0}$. Here we take $\mu=1300$~MeV and $a=-1.474$ for both channels, which are the same as those in the study of the $D_s^{*+}\rho^-$ interaction in Refs.~\cite{Molina:2020hde,Molina:2022jcd,Duan:2023qsg}. In addition, we consider the decay widths of the $\rho$ and $K^*$ mesons by means of the convolution of the two meson loop function with an energy dependent width,
	\begin{equation}
    \begin{aligned}
		\tilde{G}=&\frac{1}{N}\int^{M^2_{\rm max}}_ {M^2_{\rm min}}d\tilde{m}_1^2\left(-\frac{1}{\pi}\right)G(s^2,\tilde{m}_1^2,M_2^2) \\
     &\times {\rm Im}\left[\dfrac{1}{\tilde{m}_1^2-M_1^2+i\tilde{\Gamma}(\tilde{m}_1)\tilde{m}_1}\right],
     \end{aligned}
		\end{equation}
	with
		\begin{equation}
			N=\int^{M^2_{\rm max}}_ {M^2_{\rm min}}d\tilde{m}_1^2\left(-\frac{1}{\pi}\right){\rm Im}\left[\dfrac{1}{\tilde{m}_1^2-M_1^2+i\tilde{\Gamma}(\tilde{m}_1)\tilde{m}_1}\right],
		\end{equation}
	where $M_1$ is the nominal mass of the vector meson ($\rho$ or $K^*$), $M_{\rm min}=M_1-3.5\Gamma_0$, $M_{\rm max}=M_1+3.5\Gamma_0$~\cite{Molina:2022jcd}, with $\Gamma_0$ the resonance width at the nominal mass of the $\rho$ and $K^*$ mesons, and 
		\begin{equation}
			\tilde{\Gamma}(\tilde{m}_1)=\Gamma_0\frac{q_{\rm off}^3}{q_{\rm on}^3}\theta(\tilde{m}-m_1-m_2),
		\end{equation}
	with
		\begin{equation}
			q_{\rm off}=\dfrac{\lambda^{1/2}(\tilde{m}^2,m_1^2,m_2^2)}{2\tilde{m}},~~
			q_{\rm on}=\dfrac{\lambda^{1/2}(M_1^2,m_1^2,m_2^2)}{2M_1},
		\end{equation}
	where $m_1=m_2=m_{\pi}$ for the $\rho$, and $m_1=m_{K}$, $m_2=m_{\pi}$ for the $K^*$. 

The transition amplitudes $V'_{i\to D^0K^0}$ of Eq.~(\ref{ttcs}) are given by,
\begin{equation}\label{t21} 
	V'_{D_s^{*+}\rho^-\to D^0K^0}=\dfrac{g_{T_{c\bar{s}},D_s^{*+}\rho^-}g_{T_{c\bar{s}},D^0K^0}}{M_{D^0K^0}^2-m_{T_{c\bar{s}}}^2+im_{T_{c\bar{s}}}\Gamma_{T_{c\bar{s}}}},
\end{equation}
\begin{equation}\label{t22}
	V'_{D^{*0}K^{*0}\to D^0K^0}=\dfrac{g_{T_{c\bar{s}},D^{*0}K^{*0}}g_{T_{c\bar{s}},D^0K^0}}{M_{D^0K^0}^2-m_{T_{c\bar{s}}}^2+im_{T_{c\bar{s}}}\Gamma_{T_{c\bar{s}}}},
\end{equation}
where the $m_{T_{c\bar{s}}}=2892$~MeV and $\Gamma_{T_{c\bar{s}}}=119$~MeV are given by Ref.~\cite{LHCb:2022sfr,LHCb:2022lzp}. The constant $g_{T_{c\bar{s}},D^{*0}K^{*0}}$ corresponds to the coupling between $T_{c\bar{s}}(2900)$ and its components $D^{*0}K^{*0}$, which could be related to the binding energy by~\cite{Weinberg:1965zz,Baru:2003qq,Wu:2023fyh,Albaladejo:2022sux},
\begin{equation}\label{g_D^{*0}K^{*0}}
	g_{T_{c\bar{s}},D^{*}K^{*}}^2=16\pi(m_{D^*}+m_{K^*})^2\tilde{\lambda}^2\sqrt{\frac{2\Delta E}{\mu}},
\end{equation}
where $\tilde{\lambda}=1$ gives the probability to find the molecular component in the physical states, $\Delta E=m_{D^*}+m_{K^*}-m_{T_{c\bar{s}}}$ denotes the binding energy, and $\mu=m_{D^*}m_{K^*}/(m_{D^*}+m_{K^*})$ is the reduced mass. 

The value of the coupling constant $g_{T_{c\bar{s}},\rho^-D_s^{*+}}$ could be obtained from the partial width of $T_{c\bar{s}}(2900)\to\rho^-D_s^{*+}$, which could be expressed as follows,
\begin{equation}\label{gg2}
	\Gamma_{T_{c\bar{s}}\to \rho^-D_s^{*+}} =\frac{3}{8\pi}\frac{1}{m_{T_{c\bar{s}}}^2}|g_{T_{c\bar{s}}, \rho^-D_s^{*+}}|^2 |\vec{q}_{\rho}| ,
\end{equation}
where $\vec{q}_{\rho}$ is the three-momentum of the $\rho^-$ in the $T_{c\bar{s}}(2900)$ rest frame,
\begin{equation}
	|\vec{q}_{\rho}|=\dfrac{\lambda^{1/2}(m_{T_{c\bar{s}}}^2,m_{D_s^{*+}}^2,m_{\rho^-}^2)}{2m_{T_{c\bar{s}}}}
\end{equation}
with the K$\ddot{a}$llen function $\lambda(x,y,z)=x^2+y^2+z^2-2xy-2yz-2zx$. And similarly we can get the $g_{T_{c\bar{s}},D^0K^0}$ coupling from the partial width of $T_{c\bar{s}}(2900)\to D^0K^0$,
\begin{equation}
	\Gamma_{T_{c\bar{s}}\to D^0K^0} =\frac{1}{8\pi}\frac{1}{m_{T_{c\bar{s}}}^2}|g_{T_{c\bar{s}},D^0K^0}|^2 |\vec{q}_{K^0}| 
\end{equation}
with
\begin{equation}
	|\vec{q}_{K^0}|=\dfrac{\lambda^{1/2}(m_{T_{c\bar{s}}}^2,m_{D^{0}}^2,m_{K^0}^2)}{2m_{T_{c\bar{s}}}}.
\end{equation}
In Ref.~\cite{Yue:2022mnf}, the partial widths of decay modes $\rho^-D_s^{*+}$ and $D^0K^0$ were estimated to be ($2.96 \sim 5.3$)~MeV and ($52.6 \sim 101.7$)~MeV, respectively. In the present work, we take the center values of the estimated decay widths, i.e., $\Gamma_{T_{c\bar{s}}\to \rho^-D_s^{*+}}=4.13$~MeV and  $\Gamma_{T_{c\bar{s}}\to D^0K^0}=77.15$~MeV, and obtain the coupling constants $g_{T_{c\bar{s}},D_s^{*+}\rho^-}=2007$~MeV, $g_{T_{c\bar{s}},D^0K^0}=4697$~MeV, and $g_{T_{c\bar{s}},D^{*0}K^{*0}}=8809$~MeV, respectively. One can find that the coupling of  $g_{T^0_{c\bar{s}0}D^{*0}K^{*0}}$ is larger than two other couplings, which implies that the $D^*K^*$ component plays the dominant role for $T_{c\bar{s}}(2900)$.

\subsection{$N^*(1535)$ in the $\Lambda_b\to D^0K^0\Lambda$} \label{sec2b}

\begin{figure}[htbp]
	
	\includegraphics[scale=0.65]{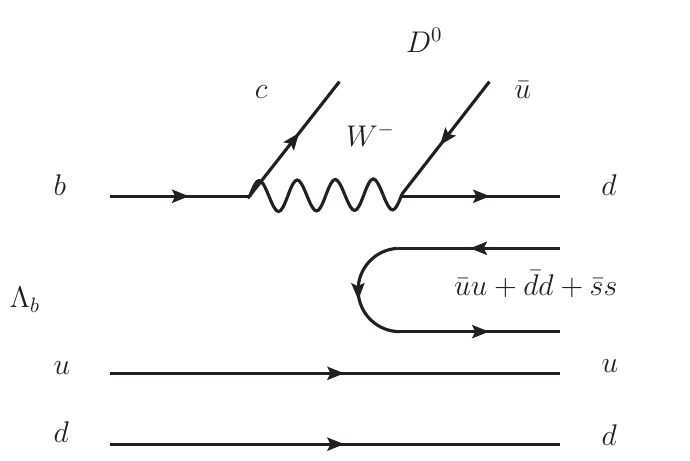}
	
	\caption{Quark level diagrams for the process $\Lambda_b\to D^0d(\bar{u}u+\bar{d}d+\bar{s}s)ud$.}\label{fig:1535-quark}
\end{figure}

In this subsection, we will present the mechanism of the process $\Lambda_b\to D^0K^0\Lambda$ via the intermediate resonance $N^*(1535)$.  As shown in Fig.~\ref{fig:1535-quark}, the $b$ quark of the initial $\Lambda_b$ weakly transits into a $c$ quark and a $W^-$ boson, then the $W^-$ boson subsequently decays into a $\bar{u}d$ quark pair. The $c$ quark from the $\Lambda_b$ decay and the $\bar{u}$ quark from the $W^-$ boson will hadronize into a $D^0$ meson. The $d$ quark from the $W^-$ boson decay and the $ud$ quark pair in the initial $\Lambda_b$, together with the quark pair $\bar{q}q=\bar{u}u+\bar{d}d+\bar{s}s$ created from the vacuum with the quantum numbers $J^{PC}=0^{++}$, hadronize into hadron pairs, which can be expressed as~\cite{Wang:2015pcn,Inoue:2001ip}, 
\begin{eqnarray}\label{eq:hadronization2}
	\left| \Lambda_b \right\rangle &=& \frac{1}{\sqrt{2}}b\left(ud-du\right) \nonumber \\
	&\Rightarrow & c W^-   \frac{1}{\sqrt{2}} \left(ud-du\right) \nonumber \\
	&\Rightarrow & c\bar{u} d \frac{1}{\sqrt{2}} \left(ud-du\right) \nonumber \\
	&\Rightarrow & D^0 d\left( \bar{u}u+ \bar{d}d +\bar{s}s\right) \frac{1}{\sqrt{2}} \left(ud-du\right) \nonumber \\
	&\Rightarrow & D^0 \left[ \pi^- \frac{u\left(ud-du\right)}{\sqrt{2}}  + \left(\frac{\eta}{\sqrt{3}} -\frac{\pi^0}{\sqrt{2}}\right) \frac{d\left(ud-du\right)}{\sqrt{2}} \right. \nonumber \\
 && \left. + K^0 \frac{s\left(ud-du\right)}{\sqrt{2}}  \right] \nonumber \\
 &=& D^0\left( \pi^-p-\frac{1}{\sqrt{2}}\pi^0n+\frac{1}{\sqrt{3}}\eta n-\sqrt{\frac{2}{3}}K^0\Lambda \right),
\end{eqnarray}
where we have the flavor wave-functions $\left|p\right>=u(ud-du)/\sqrt{2}$, $\left|n\right>=d(ud-du)/\sqrt{2}$, $\left|\pi^-\right>=d\bar{u}$, $\left|\pi^0\right>=(u\bar{u}-d\bar{d})/\sqrt{2}$, $\left|\eta\right>=(u\bar{u}+d\bar{d})/\sqrt{2}$, and $\left|K^0\right>=d\bar{s}$, and the ones of $\Lambda$ and $\Lambda_b$  are given by Eqs.~(\ref{eq:wf_lambdab}) and (\ref{eq:wf_lambda}), respectively.

\begin{figure}[htbp]
	\subfigure[]{
	\includegraphics[scale=0.65]{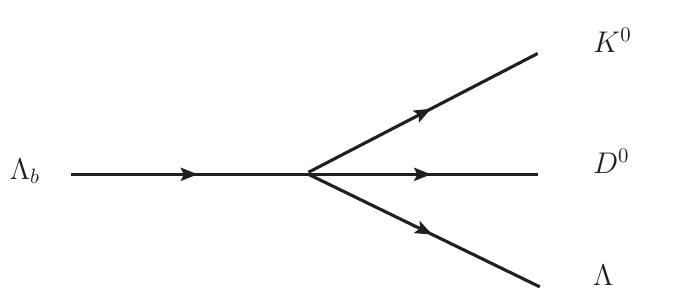}
}

\subfigure[]{
	\includegraphics[scale=0.65]{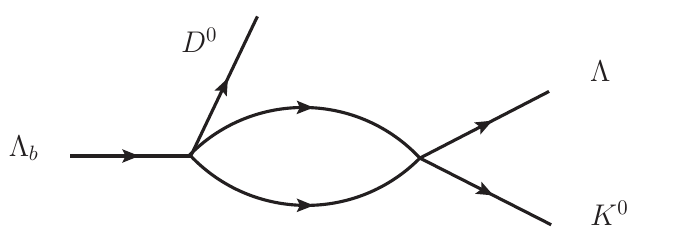}
}
	\caption{The mechanisms of the $\Lambda_b\to D^0K^0\Lambda$ decay. (a) tree diagram, (b) the $S$-wave pseudoscalar scalar-octet baryon interactions}\label{fig:1535}
\end{figure}

Then, the process $\Lambda_b\to D^0K^0\Lambda$ decay could occur through the tree diagram of Fig.~\ref{fig:1535}(a), and the $S$-wave meson-baryon interaction of Fig.~\ref{fig:1535}(b), and the amplitude could be expressed as,
\begin{equation}\label{t-N1535}
	\begin{aligned}
		\mathcal{T}^{S-wave} &= V'_{p}\left(h_{K^0\Lambda} + \sum_ih_i\hat{G}_it_{i\to K^0\Lambda}\right) ,
	\end{aligned}
\end{equation}
where the unknown constant $V'_{p}$ represents the strength of the production vertex of Fig.~\ref{fig:1535-quark} that contains all dynamical factors. The coefficients $h_i$ are obtained by Eq.~(\ref{eq:hadronization2}),
\begin{equation}\label{hi}
	\begin{aligned}
		h_{\pi^-p}=1,~h_{\pi^0n}=-\frac{1}{\sqrt{2}},~h_{\eta n}=\frac{1}{\sqrt{3}},~h_{K^0\Lambda}=-\sqrt{\frac{2}{3}}.
	\end{aligned}
\end{equation}

The $\hat{G}_i$ in Eq.~(\ref{t-N1535}) is the loop function of the meson-baryon system, and $t_{i\to K^0\Lambda}$ are the scattering matrices of the coupled channels. The transition amplitude of $t_{i\to K^0\Lambda}$ is obtained by solving the Bethe-Salpeter equation,
\begin{equation}\label{BS}
	T=[1-V\hat{G}]^{-1}V.
\end{equation}
The transition potential $V_{ij}$ is taken from Refs.~\cite{Wang:2015pcn,Inoue:2001ip}, which is
\begin{equation}\label{V}
	\begin{aligned}
		V_{ij} &=-\frac{C_{ij}}{4f_if_j}\left(2\sqrt{s}-M_i-M_j\right)\\
		&\times\left(\dfrac{M_i+E_i}{2M_i}\right)^{1/2}\left(\dfrac{M_j+E_j}{2M_j}\right)^{1/2},
	\end{aligned}
\end{equation}
where $E_i$ and $M_i$ are the energy and mass of the baryon in the $i$-th channel, and the coefficients $C_{ij}$ are shown in Table~\ref{tab:cij}, reflecting the $\mathrm{SU}(3)$ flavor symmetry. The coefficients $f_i$ are the pseudoscalar decay constants for the $i$-th channel, for which we use,
\begin{equation}
	f_{\pi}=93~\mathrm{MeV},~~~f_{K}=1.22f_{\pi},~~~f_{\eta}=1.3f_{\pi}.
\end{equation}
\begin{table}[htpb]
	\begin{center}
		\caption{ \label{tab:cij}The coefficients $C_{ij} = C_{ji}$ for the $S$-wave meson baryon scattering in the strangeness $S=0$ and isospin $I=1/2$ sector~\cite{Inoue:2001ip}.}
		\begin{tabular}{ccccccc}
			\hline\hline
			         & $K^+\Sigma^-$     &$K^0\Sigma^0$  &$K^0\Lambda$ &$\pi^-p$  &$\pi^0n$  &$\eta n$    \\
			\hline
			$K^+\Sigma^-$      & 1   &$-\sqrt{2}$ &0 &0     &$-{1}/{\sqrt{2}}$   &$-\sqrt{{3}/{2}}$ \\
						$K^0\Sigma^0$ & &0  &0  &$-{1}/{\sqrt{2}}$ &$-{1}/{2}$  &${\sqrt{3}}/{2}$ \\
			$K^0\Lambda$& & &0 &$-\sqrt{{3}/{2}}$ &${\sqrt{3}}/{2}$ &$-{3}/{2}$\\
			$\pi^-p$& & & &1&$-\sqrt{2}$&0\\
			$\pi^0n$& & & & &0 &0 \\
			$\eta n$& & & & & &0\\
				\hline\hline
		\end{tabular}
	\end{center}
\end{table}

We note that the loop function $\hat{G}_i$ function in Eq.~(\ref{t-N1535}) is calculated by using the cut off method~\cite{Oset:1997it} with $|\tilde{q}_{\rm max}|=1750$~MeV. The loop function $\hat{G}$ in Eq.~(\ref{BS}) is given by the dimensional regularization method~\cite{Wang:2015pcn,Inoue:2001ip}\footnote{It should be noted that, in a single project, it would be better to use the same regularization method to calculate these loop functions. But one can found that, in Eq.~(\ref{eq:subc}), the subtraction constants $a_i(\mu)$ for the channels of $K^0\Lambda$, $\eta n$, $\pi^-p$ and $\pi^0 n$ are positive. With a cutoff $q_{\rm max}$ in the cutoff method, the matrix $G_i$ of Eq.~(\ref{Gb}) (used by Eq.~(\ref{BS})) would imply negative subtraction constants $a_i(\mu)$, not positive ones. The need for values $a_i(\mu)>0$ is an indication that one is including the contribution of missing channels in the scattering amplitude~\cite{Hyodo:2008xr}. However, the primary $\Lambda_b\rightarrow D^0 MB$ is selective to just four channels, with particular weights, which then propagate by means of the $\hat{G}$ in Eq.~(\ref{t-N1535}) function. We are not justified to use the $\hat{G}$ function of regularization method in Eq.~(\ref{t-N1535}) of scattering to account for channels which would not contribute there~\cite{Hyodo:2008xr}.}. In this work, we take the regularization scale $\mu=1200$~MeV and use the following values for the subtraction constants $a_i$~\cite{Inoue:2001ip},
\begin{equation}
	\begin{aligned}
		&a_{K^+\Sigma^-}=-2.8,~a_{K^0\Sigma^0}=-2.8,~a_{K^0\Lambda}=1.6,\\
		&a_{\pi^-p}=2.0,~~~~~~a_{\pi^0n}=2.0,~~~~~~a_{\eta n}=0.2.
	\end{aligned} \label{eq:subc}
\end{equation}
In this work, we focus on the signal of the $T_{c\bar{s}}(2900)$ in the $DK$ invariant mass distribution, and other excited nucleons are expected to give a smooth contribution around 2.9~GeV in the $DK$ invariant mass distribution, thus we only consider the contribution from the $N(1535)$, which couples strongly to the $K\Lambda$ channel, and ignore other excited nucleons. If the experimental measurements are available in the future, one can consider the contributions from more intermediate resonances.

\subsection{Invariant Mass Distribution} \label{sec2e}

\begin{figure}
	\subfigure[]{
		\includegraphics[scale=0.65]{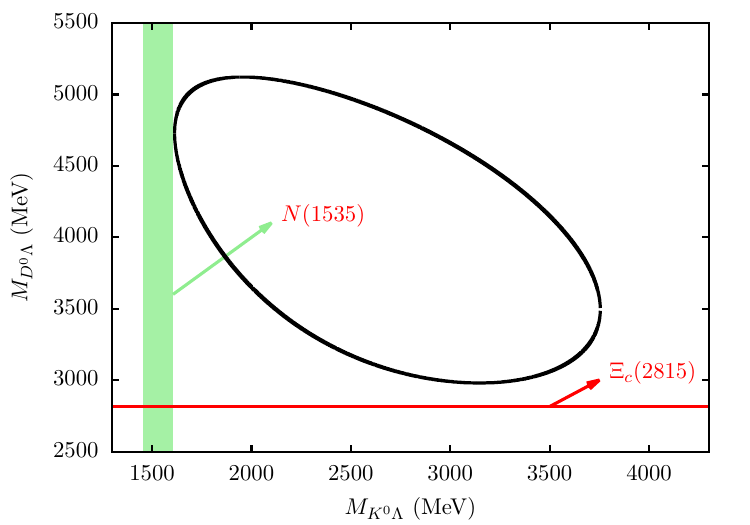}
	}
	\subfigure[]{
		\includegraphics[scale=0.65]{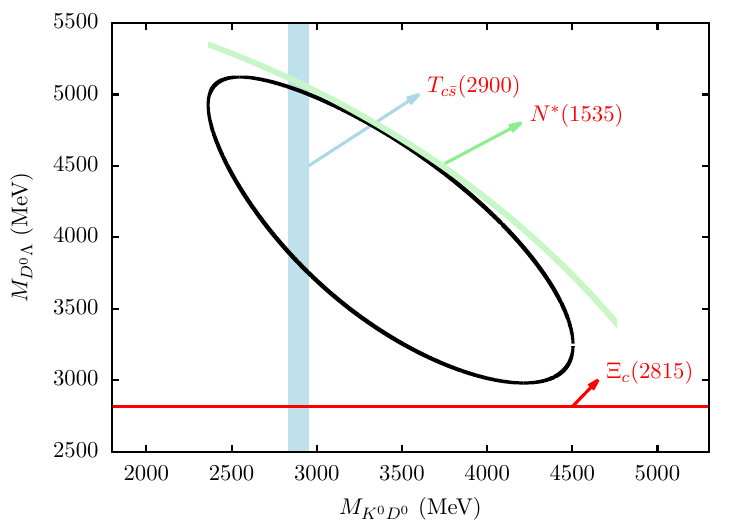}
	}
	\caption{The Dalitz plots for the process $\Lambda_b\to D^0K^0\Lambda$. The green band
		stands for the region of 1455-1605~MeV that the predicted $N^*(1535)$ state lies in, the blue band
		stands for the region of 2830-2950~MeV that the $T_{c\bar{s}}(2900)$ state lies in, while the red bands correspond to the one of $\Xi_c(2815)$, which is much narrow.}\label{fig:Dalitz}
\end{figure}

With all the ingredients obtained in the previous section, one can write down the double differential decay width for the process $\Lambda_b\to D^0K^0\Lambda$ as follows,
\begin{equation}\label{eq:dwidth}
	\dfrac{d^2\Gamma}{dM_{K^0\Lambda}dM_{D^0K^0}} = \frac{1}{(2\pi)^3}\dfrac{2M_{K^0\Lambda}2M_{D^0K^0}}{32M_{\Lambda_b}^3}|\mathcal{T}^{\text {Total}}|^2,
\end{equation}
where the total amplitude is,
\begin{equation}\label{eq:totalamp}
	\mathcal{T}^{\text {Total}}
	=\mathcal{T}^{S-wave}+\mathcal{T}^{T_{c\bar{s}}}.
\end{equation}
One could obtain the invariant mass distributions $d\Gamma/dM_{K^0\Lambda}$ and $d\Gamma/dM_{D^0K^0}$ by integrating Eq.~(\ref{eq:dwidth}) over each of the invariant mass variables. For a given value of $M_{12}$, the range of $M_{23}$ could be determined~\cite{ParticleDataGroup:2022pth}.
All the masses and widths of the particles are taken from the RPP \cite{ParticleDataGroup:2022pth}.

On the other hand, the process $\Lambda_b\to D^0K^0\Lambda$ can also occur through  the interaction of the $D^0\Lambda$, which may dynamically generate the resonances. In Ref.~\cite{Romanets:2012hm}, the interaction of the coupled channels including $D^0\Lambda$ was studied within a unitary coupled-channel approach which incorporates heavy-quark spin symmetry, and two resonances $\Xi_c(2790)$ and $\Xi_c(2815)$ are identified as the dynamically generated resonances. Since their masses are about 200~MeV below the $D^0\Lambda$ threshold, their contributions do not affect the structure of $N^*(1535)$ and $T_{c\bar{s}}(2900)$, which can be easily understood from the Dalitz plots of Figs.~\ref{fig:Dalitz}(a) and \ref{fig:Dalitz}(b). Although there are several excited $\Xi_c$ states above $D\Lambda$ threshold, it will increase the number of the free parameters if we include these intermediate $\Xi_c$ resonances. Thus, we have not taken into account their contributions, and the measurements of this process in the future could provide more information about these excited $\Xi_c$.

\section{Results and Discussions}\label{sec3}

\begin{figure}[htbp]
	
	\includegraphics[scale=0.65]{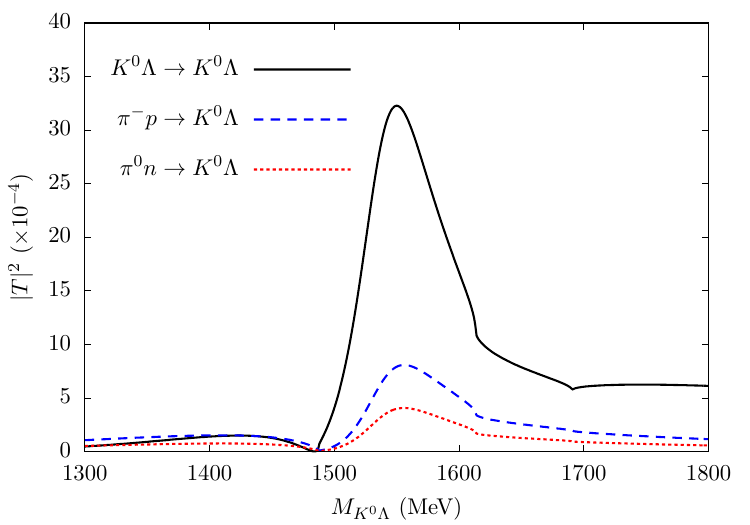}
	
	\caption{Modulus squared of the transition amplitudes $t_{i\to K^0\Lambda}$ in $S$-wave.}\label{fig:square-T}
\end{figure}

\begin{figure}
	\centering
	\includegraphics[scale=0.65]{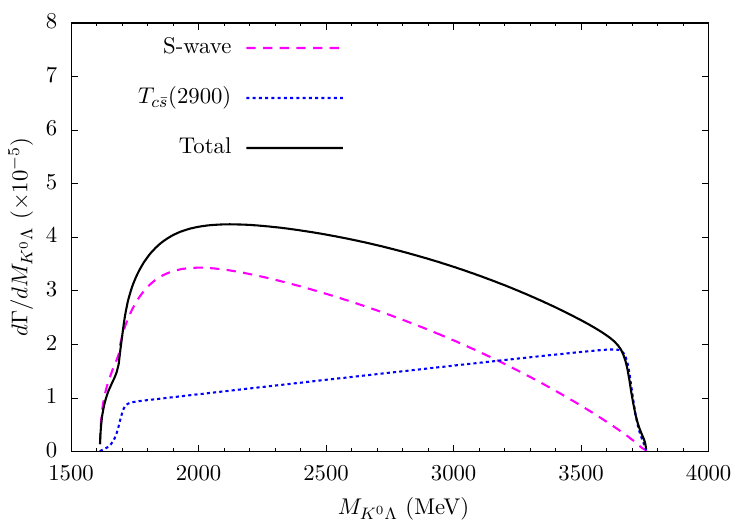}
	\caption{The $K^0\Lambda$ invariant mass distribution of the process $\Lambda_b\to D^0K^0\Lambda$ decay.}\label{fig:dwidth-KLambda}
\end{figure}

\begin{figure}
\centering
\includegraphics[scale=0.65]{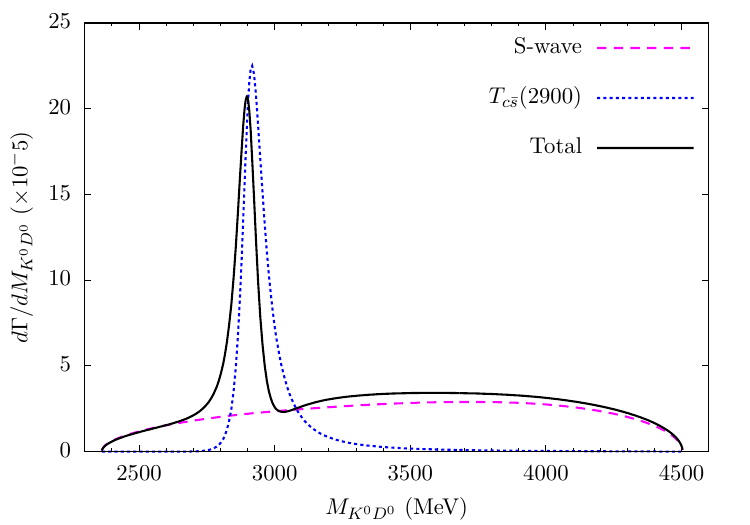}
\caption{The $D^0K^0$ invariant mass distribution of the process $\Lambda_b\to D^0K^0\Lambda$ decay.}\label{fig:dwidth-DK}
\end{figure}

\begin{figure}
\centering
\includegraphics[scale=0.80]{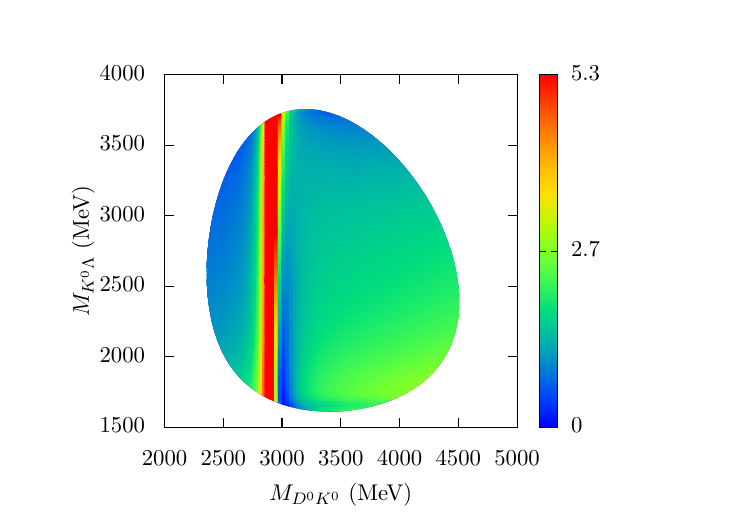}
\caption{The Dalitz plot of ``$M_{D^0K^0}$" vs. ``$M_{K^0\Lambda}$" for the double differential decay width $d\Gamma/dM_{D^0K^0}dM_{K^0\Lambda}$ the process $\Lambda_b\to D^0K^0\Lambda$. The color range is $0\sim5.3\times10^{-8}$ MeV$^{-1}$.}\label{fig:Dalize-DK-KLambda}
\end{figure}

\begin{figure*}
	\subfigure[]{
		\includegraphics[scale=0.65]{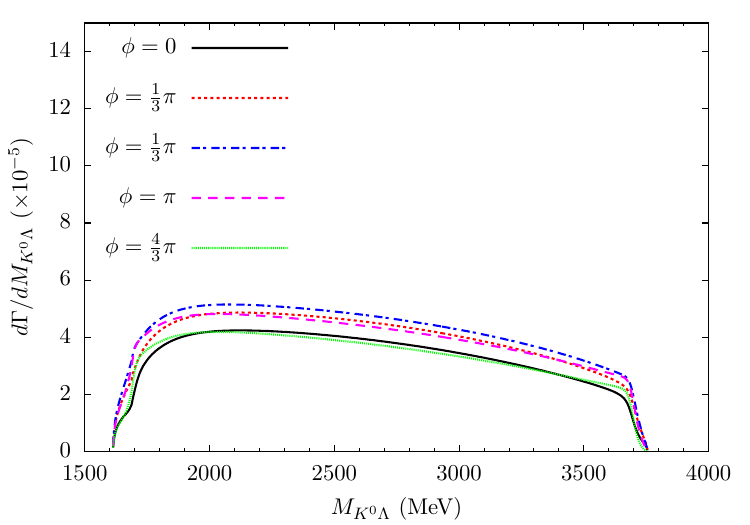}
	}
	\subfigure[]{
		\includegraphics[scale=0.65]{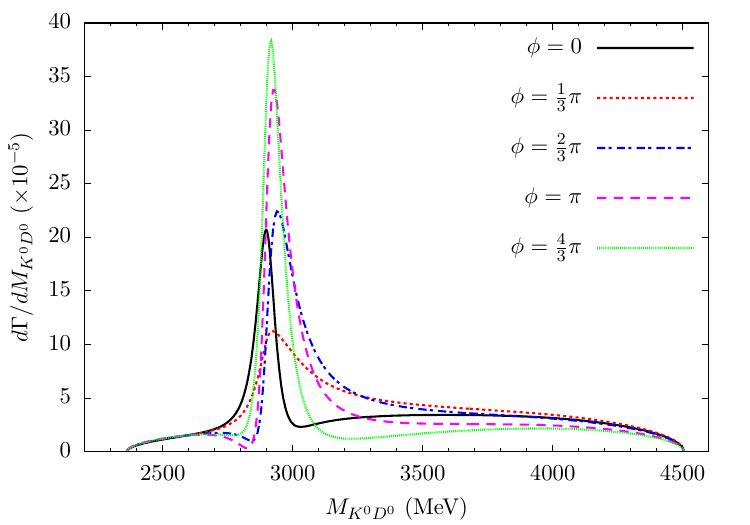}
	}
	\caption{The $K^0\Lambda$ and $D^0K^0$ invariant mass distributions of the process $\Lambda_b\to D^0K^0\Lambda$ decay with the interference phase $\phi=0$, ${\pi}/{3}$, ${2\pi}/{3}$, $\pi$, and ${4\pi}/{3}$.}\label{fig:dwidth-phi}
\end{figure*}

\begin{figure*}
	\subfigure[]{
		\includegraphics[scale=0.65]{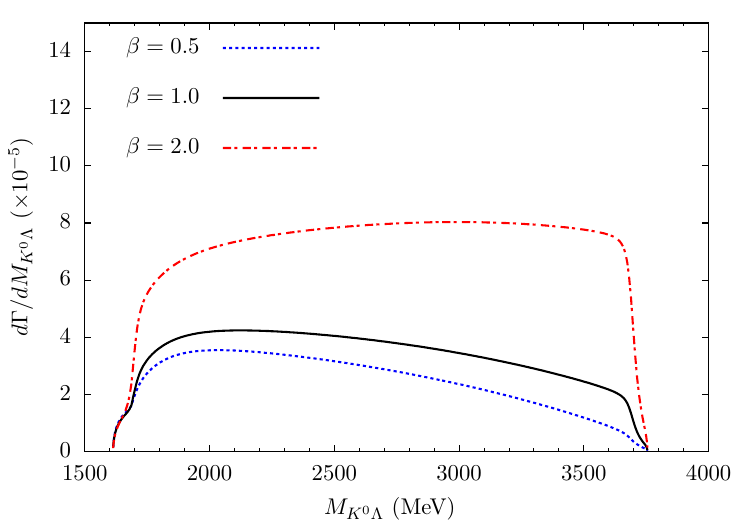}
	}
	\subfigure[]{
		\includegraphics[scale=0.65]{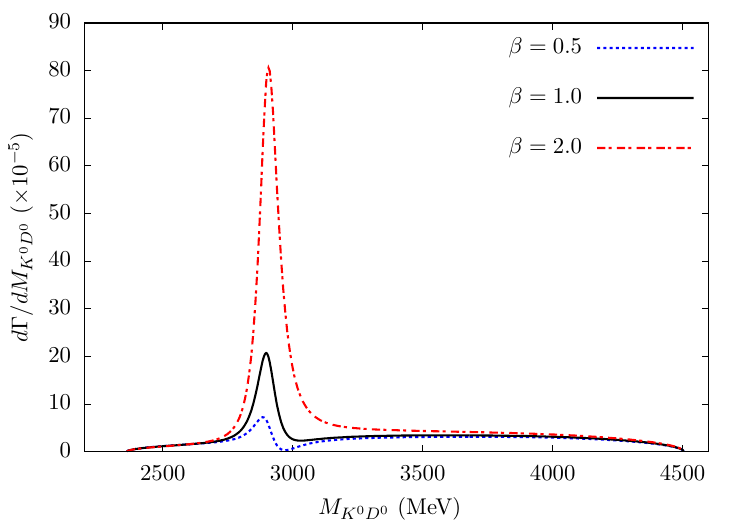}
	}
	\caption{The $K^0\Lambda$ and $D^0K^0$ invariant mass distributions of the process $\Lambda_b\to D^0K^0\Lambda$  with $\beta=V_{p}/V_{p'}=0.5$, $1.0$, and $2.0$.}\label{fig:dwidth-Vp}
\end{figure*}
Firstly, we show the modulus squared of the transition amplitudes $|t_{i\to K^0\Lambda}|^2$ of Eq.~(\ref{BS}) in Fig.~\ref{fig:square-T}. The black-solid curve shows the modulus squared of the transition amplitude $t_{K^0\Lambda \to K^0\Lambda}$, the blue-dashed curve shows the one of the transition amplitude $t_{\pi^-p\to K^0\Lambda}$, and the red-dotted curve shows the one of the transition amplitude $t_{\pi^0n\to K^0\Lambda}$. One can find that the modulus squared of the transition amplitudes $|t_{i\to D_s\Bar{K}}|^2$ exhibit a significant structure around 1535~MeV, which is associated with the resonance $N^*(1535)$~\cite{Inoue:2001ip,Wang:2015pcn}.

In our model, there are two parameters, $V_{p}$ and $V'_{p}$ corresponding to the strengths of the production vertices $\Lambda_b \to D^{*0}K^{*0}\Lambda$ and $\Lambda_b \to D^{0}K^0\Lambda$ respectively.
By now, there are no experimental information that could be used to extract the parameters $V_p$ and $V'_p$. Considering the quark level diagrams of Fig.~\ref{fig:Tcs-quark}(b) and Fig.~\ref{fig:1535-quark} are same, the parameters $V_p$ and $V'_p$ are expected to be in the same order of magnitude.
In our calculation, we assume  $V_{p}=V'_{p}$, and will discuss the uncertainties from the different ratio $\beta=V_{p}/V'_{p}$ later.   
In Figs.~\ref{fig:dwidth-KLambda} and \ref{fig:dwidth-DK}, we present our results of the $K^0\Lambda$ and $D^0K^0$ invariant mass distributions up to an arbitrary normalization, respectively. The magenta-dashed curves show the contribution from $\mathcal{T}^{S-wave}$, the blue-dotted curves show the contribution from the $\mathcal{T}^{T_{c\bar{s}}}$, and the black-solid curves show the contribution from the total amplitude. One can find a near-threshold enhancement in the $K^0\Lambda$ mass distribution, which is due to the resonance $N^*(1535)$, dynamically generated by the $S$-wave octet baryon-pseudoscalar interaction. Meanwhile, one can find a significant peak around 2900~MeV in the $D^0K^0$ invariant mass distribution, which could be associated with the tetraquark state $T_{c\bar{s}}(2900)$. 

Next, we have presented the Dalitz plot for the process $\Lambda_b\to D^0K^0\Lambda$ depending $M_{D^0K^0}$ and $M_{K^0\Lambda}$ in Fig.~\ref{fig:Dalize-DK-KLambda}. One can find that the $N^*(1535)$ mainly contributes to the low energy region of the $K^0\Lambda$ invariant mass distribution, and a clear band, corresponding to the $T_{c\bar{s}}(2900)$, appears in the region around $M_{D^0K^0}=2900$~MeV.

In addition, it should be noted that there may be the phase interference between the $\mathcal{T}^{S-wave}$ and $\mathcal{T}^{T_{c\bar{s}}}$ in Eq.~(\ref{eq:totalamp}). Thus, we multiply the amplitude $\mathcal{T}^{T_{c\bar{s}}}$ by a phase factor $e^{i\phi}$, and calculate the $K^0\Lambda$ and $D^0K^0$ invariant mass distributions with $\phi=0$, $\pi/3$, ${2\pi}/{3}$,  $\pi$, and ${4\pi}/{3}$, as presented in Figs.~\ref{fig:dwidth-phi}(a) and \ref{fig:dwidth-phi}(b). In the $K^0\Lambda$ invariant mass distribution, the near-threshold enhancement, generated by the $N^*(1535)$, appears for different values of phase $\phi$. On the other hand, although the lineshape of peak structure in the $D^0K^0$ invariant mass distribution is distorted by the interference with different phase angle $\phi$, one can always find a clear signal of the peak structure around 2900~MeV.  In the $K^0D^0$ invariant mass distribution of Fig.~\ref{fig:dwidth-phi}(b), one can find a dip structure around 3.0~GeV for phase angle $\phi=0$, which is due to the interference between different contributions~\cite{Dong:2020hxe}.

In addition, we have assumed the relationship of the parameters $V_p=V'_{p}$ in our calculations. In order to check the uncertainties of the weight parameters, we have also present the $K^0\Lambda$ and $D^0K^0$ invariant mass distributions with different values of $\beta=V_{p}/V'_{p}=0.5$, 1.0, and 2.0 in Figs.~\ref{fig:dwidth-Vp}(a) and \ref{fig:dwidth-Vp}(b), respectively. One can find that the lineshape of the $K^0\Lambda$ invariant mass distribution becomes stronger for large $\beta$, while the peak of $T_{c\bar{s}}(2900)$ in the  $D^0K^0$ invariant mass distribution is always clear. 

In this work, $T_{c\bar{s}}(2900)$ is assumed to be a $D^*K^*$ molecular state. However, if this state is assumed to be a compact tetraquark or a virtual state, the lineshape of the  $T_{c\bar{s}}(2900)$ in the $DK$ invariant mass distribution could be different. The measurements of this process in the future could provide more information about this state, which should be helpful to distinguish these different explanations.

\section{ Conclusions }

Motivated by the LHCb Collaboration measurements of the decays $B^0\to\bar{D}^0D_s^+\pi^-$ and $B^+\to D^-D_s^+\pi^+$, we have investigated the process $\Lambda_b\to D^0K^0\Lambda$ by taking into account the $D_s^{*+}\rho^-$ and $D^{*0}K^{*0}$ final-state interaction to generate the tetraquark state $T_{c\bar{s}}(2900)$. In addition, we have also taken into account the contribution from the $S$-wave pseudoscalar meson-octet baryon interaction in the $K^0\Lambda$ systems, which will dynamically generate the resonance $N^*(1535)$.

Our results show that a significant peak structure could be found around 2900~MeV in the $D^0K^0$ invariant mass distribution, which could be associated with the open-flavored tetraquark state $T_{c\bar{s}}(2900)$. Meanwhile, we have predicted  a near-threshold enhancement in the  $K^0\Lambda$ invariant mass distribution, which is due to the intermediate $N^*(1535)$. It should be pointed out that there are some unknown parameters, and we have discussed the uncertainties resulted by those parameters. The future precise measurements of the processes $\Lambda_b\to D^0K^0\Lambda$ could be used to constrain those parameters.

Since both the tree-diagram amplitudes of the processes $\Lambda_b\to D^0K^0\Lambda$ and $\Lambda_b\to D^0p\pi^-$ are in proportion to the CKM matrix elements $V_{cb}V_{ud}$, it is expected that the branching fractions of those two processes should be of the same order of magnitude, i.e. $10^{-4}$, 
 which implies that the process $\Lambda_b\to D^0K^0\Lambda$ could be accessible to be measured by LHCb in future.
On the other hand, for the $T_{c\bar{s}}(2900)^{++}$ with the quark contents $c\bar{s}\bar{d}u$, one could search for it in the $D_s^+ \pi^+$ or $D^+K^+$ mode, such as the processes $B^+\to D^+K^+\pi^-$, $B^+\to D_s^+\pi^+\pi^-$, $B^+\to D^+D^-K^+$, $B^+\to D_s^+D_s^-\pi^+$, and  $B_c^+\to D_s^+D_s^-\pi^+$.
 Therefore, we strongly encourage our experimental colleagues to measure the those processes, which would be crucial to confirm the existence the $T_{c\bar{s}}(2900)$ resonance.

\section*{Acknowledgements}
This work is supported by the National Key R$\&$D Program of China (No. 2024YFE0105200), the Natural Science Foundation of Henan under Grant No. 232300421140 and No. 222300420554,  the National Natural Science Foundation of China under Grant No. 12475086, No. 11775050, No. 12175037, No. 12335001, No. 12192263, and No. 12205075,  the China Postdoctoral Science Foundation Funded Project under Grant No. 2021M701086.

\end{document}